\documentstyle[12pt]{article}
\font\tenBbb=msbm10 \newfam\Bbbfam \textfont\Bbbfam=\tenBbb
\def\Bbb#1{{\fam\Bbbfam #1}}
\makeatletter\@addtoreset{equation}{section}\makeatother

\begin{document}
\begin{center}
{\Large LRO IN LATTICE SYSTEMS OF LINEAR CLASSICAL AND } 
\\
{\Large QUANTUM OSCILLATORS.} 
\\
{\Large STRONG N-N PAIR QUADRATIC INTERACTION.}
\end{center}
\vskip 7 pt
\begin{center}
W.I.SKRYPNIK

Intitute of Mathematics, 3 Tereshchenkivska, Kyiv 4, Ukraine, 252601.
\end{center}

\vskip 30 pt
A b s t r a c t

For systems of one-component interacting oscillators on the d-dimensional 
lattice, $d>1$, whose potential energy besides a large nearest-neighbour 
(n-n) ferromagnetic translation-invariant quadratic term contains small 
non-nearest-neighbour translation invariant term, an existence of a 
ferromagnetic long-range order for two valued lattice spins, equal to a sign 
of oscillator variables, is established  for sufficiently large magnitude 
$g$ of the n-n interaction with the help of the Peierls type contour bound. 
The Ruelle superstability bound is used for a derivation of the contour bound. 

\vskip 50 pt
\section{Introduction and main result}

Let's consider the system of one-dimensional oscillators on the $d$-
dimensional lattice $\Bbb Z^d$, with the potential energy 
(on a set $\Lambda$ with the finite cardinality $|\Lambda|$)

\begin{equation}
U(q_{\Lambda})=\sum\limits_{x\in \Lambda}(u(q_x)-2dgq_x^2)+g
\sum\limits_{|x-y|=1,x,y\in \Lambda}(q_x-q_y)^2+U'(q_{\Lambda}),
\hskip 10 pt g\geq 1
\end{equation}

Here $q_x$ the oscillator coordinate taking  value in $\Bbb R$, $q_X$ 
=$(q_x, x\in X)$, the one-particle potential  (external field) $u$ is a 
bounded below even polynomial having a degree $degu=2n$, $U'$ is an even
translation invariant function such that $U$ satisfies the superstability 
and regularity conditions, $|x|$ is the Euclidean norm of the integer valued 
vector x, $d>1$.

Let $< >_{\Lambda}, < >$ denote the Gibbs classical or quantum average
for the system confined to 
  $\Lambda$ and the system in the thermodynamic limit, i.e. $\Lambda
=\Bbb Z^d$, respectively. 

For classical systems

$$
<F_X>_{\Lambda}=
Z^{-1}_{\Lambda}\int F_X(q_X)e^{-\beta U(q_{\Lambda})}dq_{\Lambda},\hskip 15 pt
Z_{\Lambda}=\int e^{-\beta U(q_{\Lambda})}dq_{\Lambda}.
$$

If $\hat{F}_X$ is the operator of a multiplication by the function 
$F_X(q_X)$ then the quantum average is given by
$$
<F_X>_{\Lambda}=Z^{-1}_{\Lambda}Tr(\hat{F}_Xe^{-\beta H^{\Lambda}}),
\hskip 15 pt Z_{\Lambda}=Tr(e^{-\beta H^{\Lambda}})
$$
where
$H^{\Lambda}=-\frac{1}{2m}\sum\limits_{x\in \Lambda}\partial_x^2+
U(q_{\Lambda})$, and $\partial _x$ is the partial derivative in $q_x$.

The corner stone of proving an existence of lro, using generalized 
Peierls argument, is the following contour bound 

\begin{equation}
<\prod\limits_{<x,x'>\in \Gamma}\chi_x^+\chi_{x'}^->_{\Lambda} 
\leq e^{-E|\Gamma|},  
\end{equation}
where $\Gamma$ is a set of nearest neighbours, $|\Gamma|$ is the number of 
them in it, 

$$
\chi^+_x=\chi_{(0,\infty)}(q_x), \hskip 15 pt \chi^-_x=
\chi_{(-\infty,0)}(q_x),
$$
$\chi_{(a,b)}$ is the characteristic function of the open interval $(a,b)$.

The bound (1.2) was earlier derived in [BF],[FL] for several classes of
classical ferromagnetic systems or classical systems with the nearest-neighbour
pair interaction (see,also,[Si],[SH],[BW]) If one puts $s_x=sign q_x$, then taking into account
that $\chi^{+(-)}_x=\frac{1}{2}[1+(-)s_x]$ 
one obtains
$$
4<\chi^+_x\chi^-_y>_{\Lambda}=1+<s_x>_{\Lambda}-<s_y>_{\Lambda}-
<s_xs_y>_{\Lambda}.
$$

Since the systems are invariant under the transformation of changing signs 
of the oscillator variables  we have 

$$
<s_xs_y>_{\Lambda}=1-4<\chi^+_x\chi^-_y>_{\Lambda}.
$$
Now in order to prove the ferromagnetic long-range order for the spins 
$s_x$ one has to show that the average in the rhs in the equality is 
strictly less than $\frac{1}{4}$. This can be proved with the aid of the 
following lemma ([GJS], [FL])
\vskip 15 pt

LEMMA 1.1

If the bound (1.2) holds, $d>1$, and $e^{-E}$ is sufficiently small then 
there exist positive numbers $a,a'$ such that
\begin{equation}
<\chi^+_x\chi^-_y>\leq a'e^{-aE}.
\end{equation}
\vskip 15 pt
So, if one shows that  $E$ can be made arbitrary large while increasing $g$ or 
$\beta$, then the lro for the above spins will be proved.

In this paper we prove the ferromagnetic lro for
the  systems, in which interaction is neither ferromagnetic nor n-n, but 
essentially ferromagnetic for sufficiently large g (see Remark  5). We
establish (1.2) for the
simplest polynomial $u(q)=\eta q^{2n}$ with the help of the Ruelle 
superstability bound [RI] and show that  $E$ in (1,2) is positive and growing 
for growing $g$, or more precisely 

\begin{equation}
{E}=e_0-2^{-1}\ln (8\pi^{-1}e_0)-E_0, \hskip 20 pt
e_0=[g^{n}2d(\eta n)^{-1}]^{\frac{1}{2n-2}},
\end{equation}
where $E_0$ depends on $g,\beta$ and is found from 
the superstability bound (for the rescaled and translated correlation 
functions). $E_0$ is bounded in $g$ for classical systems and grows slowly 
for quantum systems (see Lemma 1.2).

The proposed technique is based on the precise knowledge how the 
constant, defining $E_0$ in the superstability bound, depends on the 
potential energy(Theorem 2.1).  It is inspired by the technique proposed
in [AKR] for quantum ferromagnetic 
systems, which by rescaling of the oscillator variables, can be reduced to the
above systems with the pair quadratic infinite-range interaction

$$
U'=\sum\limits_{x,y\in \Lambda}C_{x-y}(q_x-q_y)^2, \hskip 10 pt u(q)=\eta q^4.
$$

In this paper a complicated version of (1.2) is proposed and the lro is proved 
for Gibbs loop path system associated to the quantum system via FK formula and 
unit spin which are signs of the averaged Wiener path. A small parameter, 
appearing in the potential energy, determining a depth of the symmetric 
wells of the external potential is not associated with the magnitude 
of n-n interaction in it.

Our approach stresses the necessity to consider a large magnitude 
of n-n interaction which determines the depth of the symmetric wells of the 
external potential $u(q)-2dgq^2$ ($e_0,-e_0$ are the only its real minima). 

Proofs of an existence of an order parameter for ferromagnetic quantum
oscillator systems with n-n interaction, which are based on the 
reflection positivity, can be found in [KP],[BK] (see also [DLS]). 
Vanishing of the order parameter in the quantum limit (mass is vanishing) is 
established in [VZ].

\vskip 20 pt   

THEOREM 1.1

Let the potential energy of the one-component oscillator classical or
quantum system is given by (1.1), $u(q)=\eta q^{2n}$.
Let, also, $U'$ be a translation invariant and an even function such that
the condition of superstability and regularity hold for it 
$$
U'(q_{\Lambda})\geq -\sum\limits_{x\in \Lambda}
[Bg^{l'}|q_x|^l+B'],\hskip 10 pt 2\leq l<2n, 
$$

$$
|W'(q_{X_1;q_{X_2}})|=|U'(q_{X_1\cup X_2})-U'(q_{X_1})
-U'(q_{X_2})|\leq
$$

$$
\leq\frac{g^{l'}}{2}\sum\limits_{x\in X_1, y\in X_2}\Psi'_{|x-y|}
(|q_x|^l+|q_y|^l), \hskip 10 pt \Psi'_{|x|}\geq 0,
$$
where $l'<\frac{l}{2}, l>2;l'\leq \frac{l}{2},l=2$
for non-negative $U'$ in the second inequality and 
$l'\leq 0$ in both inequalities if $U'$ is non-positive; 
$B,B',\Psi_{|x|}$ are non-negative constants, $||\Psi'||_1=\sum\limits_{x}
\Psi'_{|x|}< \infty$ and the summation is performed over $\Bbb Z^d$. 

Then there is the feromagnetic lro in classical and quantum systems for 
the spins $s_x$ for sufficiently large $g:g>>1$, i.e $<s_xs_y> > 0$.

\vskip 20 pt 
Since $s_x$ are scale invariant and their average is not changed after 
rescaling of oscillator variables, we can deal with the rescaled by 
$g^{-\frac{1}{2}}$ variables and the potential energy $U_g$
\begin{equation}
U_g(q_{\Lambda})=\sum\limits_{x\in \Lambda}u_g^0(q_x)+
\sum\limits_{|x-y|=1,x,y\in
\Lambda}(q_x-q_y)^2+U'(g^{-\frac{1}{2}}q_{\Lambda}),
\end{equation}
where
$$
u_g^0(q)=g^{-n}\eta q^{2n}-2dq^2.
$$

The correlation functions or reduced density matrices generated by $U_g$ 
will be denoted by $\rho_g$.

The main idea of the proof originates from the inequality  
\begin{equation}
<\prod\limits_{<x,x'>\in \Gamma}\chi_x^+\chi_{x'}^->_{\Lambda} \leq
(e'e_0)^{\frac{|\Gamma|}{2}}e^{-e_0|\Gamma|}
<e^{Q_{g,\Gamma}}>_{\Lambda},
\end{equation}
where $e'$ is a positive constant, $e_0$ is a growing function of $g$, 
the expectation value is determined by $\rho_g$ and
$$
Q_{g,\Gamma}(q_{\Lambda})=
\sum\limits_{<x,x'>\in \Gamma}Q_g(q_x,q_y), \hskip 10 pt
Q_g(q_x,q_y)=\frac{1}{e_0}\{(q_x-q_{x'})^2+\frac{4}{3}(|q_x^2-
e_0^2|+|q_{x'}^2-e_0^2|)\}.
$$
Here we used the inequality
\begin{equation}
\chi^+(q_x)\chi^-(q_{x'})\leq (e'e_0)^{\frac{1}{2}}
e^{-e_0}\exp\{Q_g(q_x,q_y)]\}.
\end{equation}

Theorem 1.1 will be proved if we prove the following lemma.
\vskip 20 pt
LEMMMA 1.2

Let the conditions of Theorem 1.1 be satisfied.  Let, also, 
$e_0$ be given by (1.4). Then there exists a function $E_0(g)$ on 
the interval $[1, \infty)$ such that
\begin{equation}
<e^{ Q_{g,\Gamma}}>\leq e^{|\Gamma|E_0}.
\end{equation}

For the classical systems $E_0$ is a bounded function on the interval 
$[1,\infty)$. 

For the quantum systems if

\begin{equation}
k(g)=(1+e^{-\sqrt{\frac{g}{m}}\beta})^{-1}(1-e^{-\sqrt{\frac{g}{m}}\beta})
-\frac{20}{3}e_0^{-1}\sqrt{\frac{g}{m}}>0
\end{equation}

then there exists a bounded continuous functions 
$E_*(g)$ on $[1,\infty)$ such that  

\begin{equation}
E_0\leq \frac{1}{2}[\ln \frac{gm}{k(g)}-\ln(1-e^{-2\sqrt{\frac{g}{m}}\beta})]
+\sqrt{\frac{g}{m}}(\frac{64}{9k(g)}-\beta)+E_*(g).
\end{equation}

\vskip 20 pt
Lemmas 1.1, 1.2, i.e. (1.3) and (1.10) prove Theorem 1.1 since $e_0$ grows 
faster than $\sqrt{g}$. Function $E_*$ in the Lemma is defined by 
(3.9-11),(3.14). 
 
In the second and third sections we'll give the proof this lemma for 
classical and quantum systems, respectively.  Proofs of Lemma 1.1 
and (1.7) are standard and will not be give here (see [GJS],[FL],[AKR]).

\vskip 20 pt

\section{Lemma 1.2 via superstability argument. Classical systems.}

For classical systems with the rescaled potential energy
$$
<F_X>_{\Lambda}=
Z^{-1}_{\Lambda}\int F_X(q_X)e^{-\beta U_g(q_{\Lambda})}dq_{\Lambda}=
\int F_X(q_X)\rho_g^{\Lambda}(q_X)dq_X, 
$$
$$
\rho_g^{\Lambda}(q_X)= Z^{-1}_{\Lambda}\int
e^{-\beta U_g(q_{\Lambda})}dq_{\Lambda\backslash X},\hskip 10 pt
Z_{\Lambda}=\int\limits e^{-\beta U_g(q_{\Lambda})}dq_{\Lambda}. 
$$
Here the integration is performed over $R^{|\Lambda|}$ and $\rho^{\Lambda}$
are the correlation functions. By $\rho_g$ we'll denote the correlation
functions in the thermodynamic limit.

Changing the varibles $q_x\rightarrow q_x-e_0,$ in the integral in the 
right-hand-side of (1.6) and using the translation invariance of the 
Lebesque measure we obtain  
\begin{equation}
<e^{ Q_{g,\Gamma}}>=
\int \rho_g(q_{\Gamma}+e_0)\exp\{Q_{g,\Gamma}
(q_{\Gamma}+e_0)\}dq_{\Gamma}, \hskip 20 pt q_{\Gamma}=(q_x, q_y; <x,y>\in 
\Gamma),
\end{equation}
$$
Q_{g,\Gamma}(q_{\Gamma}+e_0)\leq\sum\limits_{<x,x'>\in 
\Gamma}\{\frac{10}{3e_0}(q_x^2+q_{x'}^2)+
\frac{8}{3}(|q_x|+|q_{x'}|)], \hskip 10 pt q_X+e_0=(q_x+e_0, x\in X).
$$
The polynomial $Q$ becomes bounded in $g$ if it is translated by $e_0$. 
As a result, we have to prove that the correlation functions,
translated by $e_0$, in the limit of growing $g$ satisfy the usual 
superstability bound. 

It is not difficult to check that if $e_0$ is given by (1.4) then

$$
u_g^0(q)=2dn^{-1}[e_0^{-2n+2}q^{2n}-nq^2].
$$

From this we immediatly see that
$$
u_g^0(q+e_0)=p_g(q)+bq^2-b',
\hskip 10 pt b=2dn^{-1}(2n(n-1)-n), \hskip 15 pt b'=2d\frac{n-1}{n}e^2_0,
$$
where $p_g$ is a bounded below polynomial in $e_0^{-1}$ and $q$ (the linear 
term  proportional to $e_0$ is absent in it)
$$
p_g(q)=2dn^{-1}\sum\limits_{s=3}^{2n}\frac{s!(2n-s)!}{n!}q^{s}e_0^{2-s}.
$$

Now we have to establish the accurate superstability and regularity conditions 
for the translated by $e_0$ potential energy.

The superstability bound is given by
\begin{equation}
U_g(q_{X}+e_0)\geq \sum\limits_{x\in X}\tilde{u}_g(q_x)-|X|B_g,\hskip 20 pt
B_g=b'+B',
\end{equation}
where 
$$
\tilde{u}_g(q)=
(u_g^0(q+e_0)+b')-Bg^{-\frac{l}{2}+l'}|q|^l.
$$
For a non-negative $U'$
$$
\tilde{u}_g(q)=u_g^0(q+e_0)+b'.
$$
\vskip 10 pt
Let's put
\begin{equation}
U_{*g}(q_{X})=U_g(q_{X}+e_0)-\sum\limits_{x\in
\Lambda}u_{*g}(q_x)+|X|B_g,
\end{equation}

$$
u_{*g}=\tilde{u}_{g}-v_g,\hskip 20 pt
v_g(q)=q^2+g^{-\frac{l}{2}+l'}|q|^l.
$$
$B_g$ diverges if $g$ tends to infinity since $b'$.  
We can add $|\Lambda|B_g$ to the potential energy since the 
expression  for the correlation functions is not changed after this. 
 
Then the following superstability condition holds

\begin{equation}
U_*(q_{X})\geq \sum\limits_{x\in X}v_g(q_x). 
\end{equation}

The regularity condition, also, holds
$$
|W_{*g}(q_{X_1};q_{X_2})|=|U_{*g}(q_{X_1\cup X_2})-U_{*g}(q_{X_1})
-U_{*g}(q_{X_2})|\leq
$$

\begin{equation}
\leq\frac{1}{2}\sum\limits_{x\in X_1, y\in X_2}\Psi_{|x-y|}
[v_g(q_x)+v_g(q_y)],\hskip 30 pt  X_1\cap X_2=\emptyset,
\end{equation} 
where $\Psi_{|x|}=2\delta_{|x|,1}+\Psi'_{|x|}$. 

Applying $|X|-1$ times the regularity condition the following important
condition is also derived

\begin{equation}
U_{*g}(q_{X})\leq
\sum\limits_{x\in X}\tilde{U}_g(q_x), \hskip 10 pt
\tilde{U}_g(q)=U_{*g}(q)+||\Psi||_1v_g(q).
\end{equation}

From the definition of the functions determining $\tilde{U}_g$, taking into 
account that $U_g(q)=u_g^0(q_)$, we derive

$$
\tilde{U}_g(q)=B'+(1+||\Psi||_1)q^2+(1+B+||\Psi||_1)g^{-\frac{1}{2}+l'}|q|^l.
$$

Let's put 
$$
\rho_{*g}^{\Lambda}(q_{\Lambda})=
\exp\{\beta\sum\limits_{x\in\Lambda}u_{*g}(q_x)\}
\rho^{\Lambda}_g(q_{\Lambda}+e_0).  
$$

Then $\rho_{*g}^{\Lambda}$ are expressed in terms of $U_{*g}$ after 
adding to $U_g$ the large in $g$ terms independent of oscillator variables

\begin{equation}
\rho^{\Lambda}_{*g}(q_X)= Z^{-1}_{*\Lambda}\int
e^{-\beta U_{*g}(q_{\Lambda})}\mu_{*g}(dq_{\Lambda\backslash X}),\hskip 10 pt
Z_{*\Lambda}=\int\limits e^{-\beta U_{*g}(q_{\Lambda})}\mu_*(dq_{\Lambda}),
\end{equation}
where
$$
\mu_{*g}(dq_Y)=\exp\{-\beta\sum\limits_{x\in Y}u_{*g}(q_x)\}dq_Y.
$$

As a result of the superstability and regularity conditions for $U_{*g}$ the 
following theorem is true [R].

\vskip 20 pt
THEOREM 2.1

Let the condition (2.4-5) hold for a positive polynomial $v_g$ and the 
function $u_{*g}$ be such that the measure $\mu_{*g}$ is finite. Then for 
arbitrary $0<3\varepsilon<1$, $r>0$ for the correlation functions defined by 
(2.7) the following (superstability) bound is valid

\begin{equation}
\rho_{*g}^{\Lambda}(q_X)\leq \exp\{-\sum\limits_{x\in X}
[\beta(1-3\varepsilon)v_g(q_x)-
c_0(I^{-1}_{r,u_{*g}},I_{u_{*g}})]\},
\end{equation}

where $c$ is a positive continuous monotonous growing at infinity  
function, 

$$
I_{r,u}=e^{-\frac{1}{2}\beta||\Psi||_1v_g(r)}I_{0u},\hskip 20 pt
I_{0u}=\int\limits_{|q|\leq r}\exp\{-\beta [\tilde{U}_{g}+
u(q)]\}dq,
$$
$$
I_u=\int \exp\{-\beta[(1-3\varepsilon)v_g(q)+u(q)]\}dq.
$$

\vskip 15 pt
We formulated the Ruelle result in a more explicit form in order to
trace the dependence in $g$ in all the terms. 

(2.1) and theorem 2.1 yields

\begin{equation}
<e^{ Q_{g,\Gamma}}>\leq e^{|\Gamma|E_0}, \hskip 10 pt
E_0=E^0+e_*(g), \hskip 10 pt e_*(g)=c_0(I^{-1}_{r,u_{*g}},I_{u_{*g}}),
\end{equation}

$$
E^0=2\ln \int \exp\{-\beta(1-3\varepsilon)v_g(q)
-\beta u_{*g}(q)+\frac{10}{3e_0}q^2+\frac{8}{3}|q|\}dq.
$$

As a result, (1.2) holds with  $E$ given by (1.4). From  the conditions 
of the Theorem 1.1 it follows that $E^0$ and $e_*$ exist in the limit 
of vanishing $g^{-1}$. Here we have to rely on the following significant 
equalities 
$$
\lim_{g^{-1}\rightarrow 0}(u_g^0(q+e_0)+b')=bq^2, \hskip 15 pt
\lim_{g^{-1}\rightarrow 0}v_g(q)=kq^2, \hskip 10 pt b\geq 4d,
$$ 
where $k=1$ or $k=2$. From the inequalities ($|q|\leq r$)

\begin{equation}
\tilde{U}_g(q)\leq
B'+(1+||\Psi||_1)r^2+(1+B+||\Psi||_1)g^{-\frac{l}{2}+l'}r^l,
\end{equation}
\begin{equation}
u_{*g}(q)\leq \tilde{u}_g(q)+v_g(q)\leq p_g(r)+br^2+v_g(r)+
Bg^{-\frac{l}{2}+l'}r^l
\end{equation}
it follows that
\begin{equation}
(I^{-1}_{0u_{*g}})^{-1}\leq r^{-1}e^{\beta p_g^+(r)},
\end{equation}

\begin{equation}
p_g^+(r)=p_g(r)+B'+(2+||\Psi||_1)r^2+(2+B+||\Psi||_1)g^{-\frac{l}{2}+l'}r^l.
\end{equation}

Polynomial $p^+_g(r)$ is uniformely bounded in $g$

\begin{equation}
p^+_g(r)\leq p^0(r)+B'+(2+||\Psi||_1)r^2+(2+B+||\Psi||_1)r^l=\bar{p}(r),
\end{equation}
$$
p^0(r)=2dn^{-1}\sum\limits_{s=3}^{2n}\frac{s!(2n-s)!}{n!}r^{s}.
$$
So, $e_*(g)$ is a bounded function.

Classical part of Lemma (1.2) is proved. Application of Lemma 1.1 completes 
the proof of Theorem 1.1 for classical systems.
\vskip 20 pt

\section{Lemma 1.2 via superstability bound. Quantum systems.}

For quantum rescaled systems  the Gibbs average of the operator $\hat{F}_X$ of
multiplication by the the function $F_X(q_X)$ is
determined by the 
reduced density matrices (RDMs) $\rho^{\Lambda}_g(q_X|q_X)$,

$$
<F_X>_{\Lambda}=Z^{-1}_{\Lambda}Tr(\hat{F}_Xe^{-\beta H_g^{\Lambda}})=
$$
\begin{equation}
=Z^{-1}_{\Lambda}(\sqrt{g})^{|\Lambda|}\int F_X(q_X)e^{-\beta
H_g^{\Lambda}}(q_{\Lambda};q_{\Lambda})dq_{\Lambda}=
\int F_X(q_X)\rho^{\Lambda}_g(q_X|q_X)dq_X,
\end{equation}

\begin{equation}
\rho^{\Lambda}_g(q_X|q_X)=(\sqrt{g})^{|X|}\int\rho^{\Lambda}_g(\omega_X)
P^{g\beta}_{q_{X},q_X}(dw_X),
\hskip 10 pt \rho^{\Lambda}_g(\omega_X)=Z_{\Lambda}^{-1}\int
e^{-U_g(\omega_{\Lambda})}P_0(d\omega_{\Lambda\backslash X}),
\end{equation}

where $\omega=(q,w)\in \Omega^*=\Bbb R\times \Omega$,  $\Omega$ is the 
probability space of Wiener paths($w\in \Omega$), $P^{t}_{q,q'}(dw)$ is the 
Wiener (conditional)measure concentrated on paths, starting from $q$ and 
arriving in $q'$ at the time $t$,
$P_0(d\omega)=\sqrt{g}dqP^{g\beta}_{q.q}(dw)$, 

$$
U_g(\omega_{\Lambda})=g^{-1}\int\limits_{0}^{g\beta}U_g(w_{\Lambda}(t))dt=
\int\limits_{0}^{\beta}U_g(w_{\Lambda}(gt))dt
$$

In deriving the formulas we applied the Feymann-Kac formula to the kernel 
\newline $e^{-\beta
H^{\Lambda}}(\sqrt{g^{-1}}q_X;\sqrt{g^{-1}}q_X')$ of the operator 
$e^{-\beta H^{\Lambda}}$and the relation

\begin{equation}
\int P^t_{\sqrt{g^{-1}}q,\sqrt{g^{-1}}q'}(dw)f(w(t_1),...,w(t_n))=
\sqrt{g}\int
P^{gt}_{q,q'}(dw)f(\sqrt{g^{-1}}w(gt_1)),...,\sqrt{g^{-1}}w(gt_n)),
\end{equation}
which follows from

$$
\exp\{t\partial^2\}(\sqrt{g^{-1}}q;\sqrt{g^{-1}}q')=(4\pi
t)^{-\frac{1}{2}}\exp\{-\frac{|q-q'|^2}{4tg}\}=
\sqrt{g}\exp\{tg\partial^2\}(q;q').
$$

The rescaled Hamiltonian is given by 

$$
H_g^{\Lambda}=g(-\frac{1}{2m}\sum\limits_{x\in \Lambda}\partial_x^2+
g^{-1}U_g(q_{\Lambda}))
$$

In order to prove Lemma 1.2 one has to estimate
$\rho^{\Lambda}_g(q_X+e_0|q_X+e_0)$. 

From the translation invariance of the conditional Wiener measure and the 
measure $P_0$ it follows that 

$$
\rho^{\Lambda}_g(q_X+e_0|q_X+e_0)=(\sqrt{g})^{|X|}
\int\rho^{\Lambda}_g(\omega_X+e_0)P^{g\beta}_{q_{X},q_X}(dw_X),
$$

$$
\rho^{\Lambda}_g(\omega_X+e_0)=Z_{\Lambda}^{-1}\int
e^{-U_g(\omega_{\Lambda}+e_0)}P_0(d\omega_{\Lambda\backslash X}), \hskip 10 pt
Z_{\Lambda}=\int e^{-U_g(\omega_{\Lambda}+e_0)}P_0(d\omega_{\Lambda}),
$$
where $\omega+e_0=w(t)+e_0, w(0)=q, t\in [0,t]$.

As a result    
$$
<e^{ Q_{g,\Gamma}}>_{\Lambda}=(\sqrt{g})^{|\Gamma|}\int
e^{Q_{g, \Gamma}(q_{\Gamma}+e_0})
\rho^{\Lambda}_g(\omega_X+e_0)dq_{\Gamma}
P^{g\beta}_{q_{\Gamma},q_{\Gamma}}(dw_{\Gamma}),
$$
It is evident that Lemma 1.2 can be proved now with 
the help of the analogue of the superstability bound for
$\rho_g^{\Lambda}(\omega_{\Lambda}+e_0)$ which was proved by Park [P]. 

In order to prove the analogue of Theorem 2,1 one has to derive the
superstability and regularity conditions for 
$U_{*g}(\omega_{\Lambda})=U_g(\omega_{\Lambda}+e_0)$. But 
now it is easy since in the previous section we estalished them for 
$U_g(q_{\Lambda}+e_0)$.

So, let by $u_{*g}(\omega), v_g(\omega), \tilde{U}_g(\omega)$
 be denoted the corresponding
functions, depending on $w(gt)$, being integrated by dt on the interval 
$[0,\beta]$ and 

$$
U_{*g}(\omega_{\Lambda})=\int\limits_{0}^{\beta}U_{*g}(w_{\Lambda}(gt))dt.
$$
where $U_{*g}(w_{\Lambda}(gt))$ is defined by (2.3)(instead of $\omega$ $w$ may be written).
Then

\begin{equation}
U_*(\omega_{X})\geq \sum\limits_{x\in X}v_g(\omega_x). 
\end{equation}

$$
|W_{*g}(\omega_{X_1};\omega_{X_2})|=|U_{*g}(\omega_{X_1\cup
X_2})-U_{*g}(\omega_{X_1})-U_{*g}(q_{X_2})|\leq
$$

\begin{equation}
\leq\frac{1}{2}\sum\limits_{x\in X_1, y\in X_2}\Psi_{|x-y|}
[v_g(\omega_x)+v_g(\omega_y)],\hskip 30 pt  X_1\cap X_2=\emptyset,
\end{equation} 

$$
U_*(\omega_{X})\leq \sum\limits_{x\in X}\tilde{U}_g(\omega_x).
$$

Let's put 
$$
\rho_{*g}^{\Lambda}(\omega_{\Lambda})=
\exp\{\sum\limits_{x\in\Lambda}u_{*g}(\omega_x)\}
\rho^{\Lambda}_g(\omega_{\Lambda}+e_0).  
$$

Hence

\begin{equation}
\rho^{\Lambda}_{*g}(\omega_X)= Z^{-1}_{*\Lambda}\int
e^{-U_{*g}(\omega_{\Lambda})}P_{*0}(d\omega_{\Lambda\backslash
X}),\hskip 10 pt
Z_{*\Lambda}=\int e^{-U_{*g}(\omega_{\Lambda})}P_{*0}(d\omega_{\Lambda}),
\end{equation}
where
$$
P_{*0}(d\omega_Y)=\exp\{-\sum\limits_{x\in Y}u_{*g}(\omega_x)\}
P_{0}(d\omega_Y).
$$

As a result of (3.3-5) the following theorem is true.
\vskip 20 pt

THEOREM 3.1

Let the condition (3.4-5) hold for a positive polynomial $v_g(q)$ and 
the functional $u_{*g}$ be
such that the measure $P_{*0}$ is finite. Then for arbitrary 
$0<3\varepsilon<1$, $r>0$ for the correlation functions defined by (3.6) 
the following (superstability) bound is valid

\begin{equation}
\rho_{*g}^{\Lambda}(\omega_X)\leq \exp\{-\sum\limits_{x\in X}
[(1-3\varepsilon)v_g(\omega_x)-c_0(I^{-1}_{r,u_{*g}},I_{u_{*g}}))]\},
\end{equation}
where is a 
positive continuous monotonous growing at infinity  function,
$$
I_{r,u}=e^{-\frac{1}{2}\beta ||\Psi||_1\bar{v}_{g,r}}
I_{0u},\hskip 20 pt \bar{v}_{g,r}=ess\sup\limits_{\omega
\in \Omega_r^*}v_g(\omega),
$$
$$
I_{0u_{*g}}=\int\limits_{\Omega_{r}^*}\exp\{-[\tilde{U}_g(\omega)+
u_{*g}(\omega)]\}P_{0}(d\omega),\hskip 20 pt  
I_{u_{*g}}=\int
\exp\{-[(1-3\varepsilon)v_g(\omega)+u_{*g}(\omega)]\}P_{0}(d\omega),
$$
where $\Omega_{r}^*=\{\omega\in \Omega^*:|w(t)|\leq r\}$.
\vskip 15 pt
The proof of this theorem does not essentially differs from the proof of 
Theorem 2.1.

Proof of Lemma 2.1.

Theorem 3.1 yields the following equalities

\begin{equation}
<e^{ Q_{g,\Gamma}}>\leq e^{|\Gamma|E_0}, \hskip 10 pt
E_0=E^0+e_*(g), \hskip 10 pt e_*(g)=c_0(I^{-1}_{r,u_{*g}},I_{u_{*g}}),
\end{equation}

$$
E^0=2\ln \sqrt{g}\int \exp\{-\beta[(1-3\varepsilon)v_g(w)-u_{*g}(w)]
+\frac{10}{3e_0}q^2+\frac{8}{3}|q|\}P_{q.q}^{g\beta}(dw)dq.
$$
$e_*(g)$ is a bounded in $g$ since $\tilde{U}_g(w), u_{*g}(w)$ are 
polynomial functionals and the function $c$ is continuous. 

For $E^0$ the following bound is valid after adding to the argument of the 
exponent $\frac{1}{4} (w_g^{,2}(\beta)-w_g^{,2}(\beta))$ and applying the 
Schwartz inequality ( for the measure $\sqrt{g}dqP^{g\beta}_{(q,q)}(dw)$) 

$$
e^{E^0}\leq gI^0I_0,
$$

$$
I^0=\int \exp \{\frac{20}{3e_0}q^2+\frac{16}{3}|q|-
\frac{1}{2}w^{,2}_g(\beta)\}P_{q.q}^{g\beta}(dw)dq,
$$

$$
I_0=\int \exp\{-2\beta[(1-3\varepsilon)v_g(w)+u_{*g}(w)]+\frac{1}{2}
w_g^{,2}(\beta)\}P_{q.q}^{g\beta}(dw)dq,
$$
where $w_g^{,2}(\beta)=\int\limits_{0}^{\beta}w^2(gt)dt$. 

From the FK formula it follows that $I_0$ is the trace of the kernel of the 
exponent of perturbed generator of the Wiener process. So, the Golden-
Thompson inequality $Tr(e^{A+B})\leq Tr(e^Ae^B)$ yields 

\begin{equation}
gI_0\leq \sqrt{gm}
(2\pi \beta)^{-\frac{1}{2}}
\int \exp\{-2\beta[(1-3\varepsilon)v_g(q)+u_{*g}(q)+\frac{q^2}{4}]\}dq=
\sqrt{gm}I_0^-.
\end{equation}
Here we took into account that 
$$
\exp\{t\partial^2\}(q;q')=(4\pi t)^{-\frac{1}{2}}\exp\{-\frac{|q-q'|^2}{4t}\}.
$$
$I_0$ is finite since $b\geq 4d$.

For $I^0$ after the rescaling $q=(m^{-1}g)^{\frac{1}{4}}\tilde{q}$ we have
($\hat{q}^2$ is the operator of multiplication by $q^2$)
$$
I^0=(\frac{g}{m})^{\frac{1}{4}} \int 
\exp\{-g\beta(-(2m)^{-1}\partial^2+\frac{1}{2g}\hat{q}^2)\}((m^{-1}g)^{\frac{1}{4}}q,
(m^{-1}g)^{\frac{1}{4}}q)\times
$$
$$
\times\exp\{3^{-1}(20e_0^{-1}\sqrt{m^{-1}g}q^2+
8(m^{-1}g)^{\frac{1}{4}}|q|)\}dq.
$$
From (3.3) it follows that
$$
(\frac{g}{m})^{\frac{1}{4}}\exp \{-g\beta(-\frac{1}{2m}\partial^2+\frac{1}{2g}\hat{q}^2)\}
((m^{-1}g)^{\frac{1}{4}}q,(m^{-1}g)^{\frac{1}{4}}q')=
$$
$$
=e^{-2^{-1}\sqrt{\frac{g}{m}}\beta}\exp\{-\sqrt{\frac{g}{m}}\frac{\beta}{2}
(-\partial^2+\hat{q}^2-1)\}(q,q')=
$$
$$
=e^{-2^{-1}\sqrt{\frac{g}{m}}\beta}\exp\{-\frac{q^2}{2}+\frac{q'^2}{2}\}
\exp \{-\sqrt{\frac{g}{m}}\beta
(-\frac{1}{2}\partial^2+\hat{q}\partial)\}(q,q')=
$$
$$
=\sqrt{\pi^{-1}}(1-e^{-2\sqrt{\frac{g}{m}}\beta})^{-\frac{1}{2}}
\exp\{ -\frac{q^2}{2}+
\frac{q^{'2}}{2}-(1-e^{-2\sqrt{\frac{g}{m}}\beta})^{-1}
(q'-e^{-\sqrt{\frac{g}{m}}\beta}q)^2-2^{-1}\sqrt{\frac{g}{m}}\beta\}.
$$
Here we used the relation

$$
\frac{1}{2}(-\partial^2+\hat{q}^2-1)=
e^{-\frac{\hat{q}^2}{2}}
[-\frac{1}{2}\partial^2+\hat{q}\partial]e^{\frac{\hat{q}^2}{2}}
$$
and the well-known formula for the density of the transition 
probability for the Ornstein-Uhlunbeck process. Hence

$$
I^0=\sqrt{\pi^{-1}}e^{-\sqrt{\frac{g}{m}}\beta}
(1-e^{-2\sqrt{\frac{g}{m}}\beta})^{-\frac{1}{2}}
\int \exp\{3^{-1}(20e_0^{-1}\sqrt{\frac{g}{m}}q^2+
8(\frac{g}{m})^{\frac{1}{4}}|q|)\}\times 
$$
$$
\times\exp\{-
(1+e^{-\sqrt{\frac{g}{m}}\beta})^{-1}
(1-e^{-\sqrt{\frac{g}{m}}})q^2\}dq.
$$
As a result 

\begin{equation}
I^0=
(1-e^{-2\sqrt{\frac{g}{m}}\beta})^{-\frac{1}{2}}k(g)^{-\frac{1}{2}}
\exp\{\frac{64}{9} k(g)^{-1}\sqrt{\frac{g}{m}}\},
\end{equation}

\begin{equation}
e^{E^0}\leq \sqrt{mg}e^{-\sqrt{\frac{g}{m}}
(\beta-\frac{64}{9}k(g)^{-1})}
(1-e^{-2\sqrt{\frac{g}{m}}\beta})^{-\frac{1}{2}}k(g)^{-\frac{1}{2}}I_0^- , 
\end{equation}
where $k(g)$ is given by (1.9).

Applying the Golden-Thompson inequality we obtain 

\begin{equation}
I_{u_{*g}} \leq \sqrt{m}I_0^-.
\end{equation}

Repeating (2.10-11), using the equality 
$$
\sqrt{g}\int P_{q,q}^{g\beta}(dw)=\sqrt{(2\pi \beta)^{-1}m}, 
$$
we derive, also, the analogue of (2.12) for the quantum case

\begin{equation}
(I^{-1}_{0u_{*g}})^{-1}\leq \sqrt{2\pi \beta m^{-1}} r^{-1}e^{\beta p^+(r)}.
\end{equation} 
As a result

Combining all these bounds we see that $e_{*}(g)$ is bounded and (1.10) holds
with

\begin{equation}
E_*(g)=\ln I_0^-+e_{*}(g).
\end{equation}

Lemma 1.2 is proved. Theorem 1.1 is proved with an aid of Lemma 1.1 and (1.7).

\vskip 10 pt
REMARKS. 

1.If one cancells the boundary term 
$g\sum\limits_{x\in \partial \Lambda}q_x^2$  then (1.1) is reduced to

$$
U(q_{\Lambda})=\sum\limits_{x\in \Lambda}u(q_x)-g
\sum\limits_{|x-y|=1,x,y\in \Lambda}q_xq_y+U'(q_{\Lambda}).
$$
where $\partial X$ is the boundary of $X$. 
If $U'>0$ then the systems considered in [BF] can be recovered

Surprisingly the proposed technique does not respect this boundary term 
since it creates an obstruction for obtaining the bound from above  
(2,6) for the rescaled and translated potential energy which guarantees 
uniform boundedness in $g$ of $I_{0u_{*g}}^{-1}$.

2. If one cancells the boundary term $\sum\limits_{x\in \partial X}u(q_x,q_x)$
then (1.1) is equal to
$$
U(q_X)=\sum\limits_{x,y\in
X,|x-y|=1}u(q_x,q_y)+U'(q_X)\hskip 10 pt
u(q_x,q_y)=(4d)^{-1}(u(q_x)+u(q_y))-gq_xq_y
$$

If $U'$ is expressed in the same form as the first term in the rhs of the last 
equality then systems which are dealt with in [FL] can be recovered. Theorem 
(1.1) can be proved for such the potential energy taking into account  
in a special way a contribution of the boundary term to the superstability, 
regularity conditions and (2.6) for the rescaled and translated potential 
energy(see [S]).

3.Theorem 1.1 proves an existence of a phase transition for the case $U'$ 
is expressed through a pair (special)potential, since it is known that in 
this case in the high-temperature phase there is an exponential decrease 
of correlations [K].

4. The magnitude of n-n interaction plays an exceptional role in the proposed 
approach since vanishing of it automatically implies vanishing of the spin 
two-point function for n-n sites. This means that $E$ in (1.2) has 
to depend on the magnitude of n-n interaction, tending to zero 
together with it. So, one ought, always, to rescale by the magnitude 
 (in an appropriate power) all the variables, when starting to derive the 
Peierls type contour bound using (1.7) with $e_0$ depending on it. 
   
5. Essentially ferromagnetic interaction may be characterized by the property 
that the ferromagnetic configuration, consisting of the coordinate $e_0$
(minimum of a one-particle potential) at 
each lattice site, is more favorable than the associated antiferromagnetic 
(staggered) configuration, consisting of the coordinate $e_0$ at the even 
sublattice and $-e_0$ at the odd sublattice for sufficiently large $g$, 
i.e. the potential energy on the former configuration is less than on the 
latter. This property follows from the superstability condition for the 
rescaled $U'_g$ in the formulation of Theorem 1.1 and the fact that the 
growth in $g$ of $g^{-s}e_0^{2s}, s<n$, is more slow than $e_0^2$. In other 
words, the ferromagnetic n-n part of the potential energy suppresses 
antiferromagnetic ground states for sufficiently large $g$.

6. If one puts

$$
U'(q_{\Lambda})=\sum\limits_{x,y\in \Lambda}C_{x-y}(q_x-q_y)^2, \hskip 10 pt
|C_{x-y}|\leq C^0_{|x-y|}, \hskip 10 pt ||C^0||_1<\infty,
$$
where $||C^0||_1$ does not depend 0n $g$ then the conditions of Theorem 1.1 
are satisfied. 

7. The  proof of Theorem 1.1 for classical systems and more general polynomial 
potentials $u$ can be found in [S]
\vskip 20 pt 

AKNOWLEDGMENTS.

The research described in this publication was possible in part by the Award 
number UP1-309 of the U.S. Civilian Research and Development Foundation 
(CRDF) for independent states of the former Soviet Union.

The author expresses gratitude to prof. V.Priezzhev for discussions.

\vskip 20 pt
REFERENCES

[AKR] S.Albeverio, O.Kondratiev, O.Rebenko, Journ.Stat.Phys., 92, 5/6, p.1137, 
1998. 

[BF] J.Bricmont, J.-R.Fontaine, Journ.Stat.Phys., 26, N4, p.745, 1981.

[BK] V.Barbulyak, Yu.Kondratiev, Reports Nat.Acad.Sci. of Ukraine, 10, p.19, 
1991.

[BW] C.Borgs, R.Waxler, Commun.math.phys.,126, p.683, 1990.

[DLS] F.Dyson, E.Lieb, B.Simon, J.Stat.Phys., 18, p.335,1978.

[FL] J.Frohlich, E.Lieb, Comm.Math.Phys., 60, p.233, 1978.

[GJS] J.Glimm, A.Jaffe, T.Spencer, Commun.Math.Phys.,45, p.203.(1975).
 
[KP] B.Khoruzhenko, L.Pastur,  Teor.Mat.Fiz.,73, p.1094, 1987.

[P] Y.Park, J.Korean Math.Soc., 23, p.43, 1985.

[K] H.Kunz, Comm.math.phys.,59, p.53, 1978.

[R] D.Ruelle,  Comm.Math.phys., 18,p,127, 1970.

[Sh] S.Shlosman, Uspehi Math.Sci., v.41, N3(249), p.69, 1986 (in Russian).

[Si] Ya.G. Sinai, Phase transitions. Rigorous results. Moscow. Nauka. 1980
(there is the English translation).

[S] W.Skrypnik, J.Phys.A.(to be published)

[VZ] A.Verbeure, V.Zagrebnov, No-go theorem for quantum structural phase 
transitions. Preprint-KUL-TP-95/05.

\end{document}